\documentclass{aa}  

\usepackage{graphicx}
\usepackage{ulem}
\usepackage{txfonts}
\begin{document}

   \title{An estimate of the $k_{2}$ Love number of WASP-18Ab from its radial velocity measurements}


   \author{Sz. Csizmadia
          \and
          H. Hellard
          \and
          A. M. S. Smith
          }

   \institute{Deutsches Zentrum f\"ur Luft- und Raumfahrt, Institut f\"ur Planetenforschung,
              Berlin, Rutherfordstr 2, D-12489, Germany\\
              \email{szilard.csizmadia@dlr.de}
             }

   \date{Received 4 October 2018; accepted 29 Nov 2018}

 
  \abstract
   {Increasing our knowledge of the interior structure, composition and density    
    distribution of exoplanets is crucial to make progress in the understanding of
    exoplanetary formation, migration and habitability. However, the directly 
    measurable mass and radius values offer little constraint on interior structure, 
    because the inverse problem is highly degenerate. Therefore there is a clear need for a 
    third observable of exoplanet interiors. This third observable can be the $k_2$
    fluid Love number which measures the central mass concentration of 
    an exoplanet.}
   {The aims of this paper are (i) to develop a basic model to fit the long-term radial velocity and TTV 
    variations caused by tidal interactions, (ii) to apply the model to the WASP-18Ab system, and (iii) to 
    estimate the Love number of the planet.}
   {Archival radial velocity, transit and occultation timing data are collected and fitted via the model introduced here.}
   {The best model fit to the archival radial velocity and timing data of WASP-18Ab 
   was obtained with a Love number of the massive ($\sim 10 M_\mathrm{Jup}$) hot Jupiter 
   WASP-18Ab: $k_{2,Love} = 0.62^{+0.55}_{-0.19}$. This causes apsidal motion in the 
   system, at a rate of $\sim0.0087\pm0.0033^\circ / \mathrm{days} \approxeq 
   31.3\pm11.8$ arcseconds/day. When checking possible causes of 
   periastron precession, other than the relativistic term or the non-spherical shape of the components, 
   we found a companion star to the WASP-18 system, named WASP-18B 
   which is a probable M6.5V dwarf with $\sim 0.1~M_\odot$ at  3519 AU distance from the 
   transit host star. We also find that small orbital eccentricities may be real, rather than an apparent effect caused by the non-spherical stellar shape.}
  {}

   \keywords{ Techniques: radial velocities -- 
              Planets and satellites: interiors --
              Planets and satellites: individual: WASP$-18$Ab --
              Methods: data analysis
            }

   \maketitle
%

\section{Introduction}

In a binary system consisting of either two stars, or a star and a planet, tidal forces act on the non-mass point component(s). These tidal forces lead to several phenomena, including a decrease of the semi-major axis (tidal decay), a decrease in the orbital eccentricity (circularization), a decrease of the rotational rate (synchronization), and a precession of the orbit in eccentric cases, i.e. the semi-major axis of the orbit rotates. This latter phenomenon is called apsidal motion or perihelium/periastron precession. Its time-scale is much shorter (10-1000 years) than the time-scale of the other listed effects (10-1000 Myrs). The periastron point of an exoplanet's orbit is at an angle $\omega$ from the tangential line of the sky (Fig. 1.). We point out here that if apsidal motion is present then this angle is time-variable, resulting in transit timing variations (TTVs) as well as radial velocity (RV) variations (see also Ferrero et al. 2013, Schmitt et al. 2016, Rauw et al. 2016) and its rate is governed by the Love numbers.

The fluid Love numbers $k_{i,Love}$ ($i=2,3,4...$) of celestial bodies measure the reaction of a body to perturbing forces. Assuming the body is in hydrostatic equilibrium, they are a direct, but complicated function of the internal radial density distribution. In the simplest two-layer, core-mantle models with polytropic equation of state, $k_{2,Love} \sim f(M_{core} / M_{total})$ where $M_{core}$ and $M_{total}$ are the core and total masses of the planet, respectively (Becker \& Batygin 2013). The full derivation of the Love numbers depend on the exact internal density-profile and is given in e.g. Padovan et al. (2018) Kellermann et al. (2018), Kramm et al. (2011), Kopal (1959) and Love (1911). The goal of this paper is to develop a model that includes the apsidal motion effect in the study of RV and TTV data and to fit this model to the available WASP-18 data, to estimate the planetary fluid Love number. Of course the presence of apsidal motion requires an eccentric orbit, and the non-circular nature of the WASP-18Ab orbit is discussed in Section 6.2.


\section{The causes of apsidal motion}

The argument of periastron changes because of general relativistic effects at a rate of (Einstein 1915):
\begin{equation}
\dot{\omega}_{GR} = \frac{6 \pi GM_{star}}{a c^2 (1-e^2)}
\end{equation}
and as a result of the non-sphericity and rotation of the components (star and planet), which produce rotational potential changes and tidal interaction (Sterne 1939):
\begin{eqnarray}
\dot{\omega}_N  & = & n k_{2,planet} \left( \frac{R_{planet}}{a} \right)^5 \frac{(P_{orb}/P_{rot,planet})^2}{ (1-e^2)^2} \left( 1  + \frac{M_{star}}{M_{planet}} \right) \nonumber \\
&+& n k_{2,planet} \left( \frac{R_{planet}}{a} \right)^5 \frac{15 M_{star}}{M_{planet}} \frac{1 + \frac{3}{2}e^2 + \frac{1}{8}e^4}{(1-e^2)^5}  \nonumber \\
& + & n k_{2,star} \left( \frac{R_{star}}{a} \right)^5 \frac{(P_{orb}/P_{rot,star})^2}{ (1-e^2)^2} \left( 1  + \frac{M_{planet}}{M_{star}} \right) \nonumber \\
&+& n k_{2,star} \left( \frac{R_{star}}{a} \right)^5 \frac{15 M_{planet}}{M_{star}} \frac{1 + \frac{3}{2}e^2 + \frac{1}{8}e^4}{(1-e^2)^5}  \nonumber \\
\end{eqnarray}
Here $N$ denotes the "Newtonian"-term, $n$ is the mean motion, $2k_2$ is the 2nd order fluid Love number, $a$ is the semi-major axis, $e$ is the eccentricity, $R_{star}$ and $R_{planet}$ are the radii of the star and planet, respectively and $M_{star}$ and $M_{planet}$ are the masses of the star and planet, respectively. $P_{orb}$, $P_{rot,star}$, and $P_{rot,planet}$ are the orbital period and the rotational periods of the star and planet, respectively.

%
   \begin{figure}
   \centering
   \includegraphics[width=8cm]{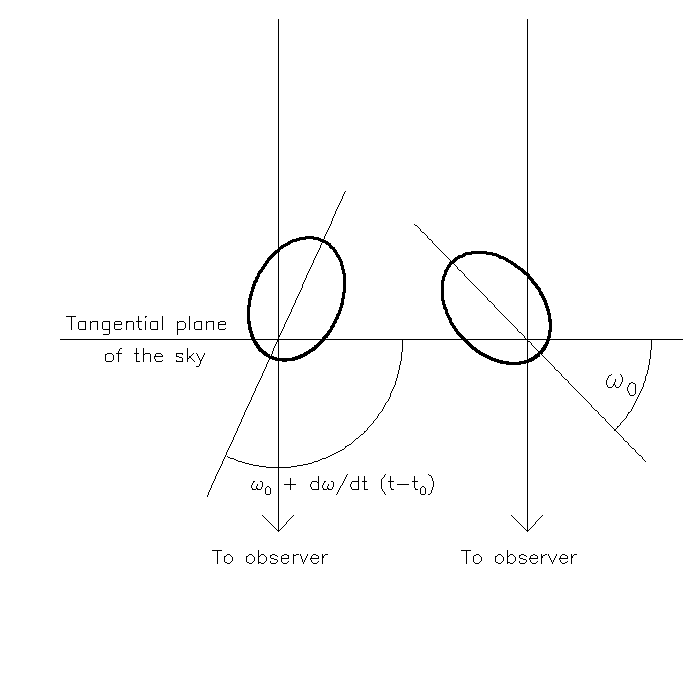}
      \caption{Illustration of the apsidal motion. At time $t_0$ the apsis 
              (semi-major axis of the orbit) is oriented at an angle of $\omega_0$ 
              (right). At 
              time $t$ it moved to the direction characterized by $\omega = \omega_0 + 
              \frac{d\omega}{dt} (t-t_0)$ (left). Traditionally, the argument of periastron is 
              measured from East to the direction of the observer, however, the radial 
              velocity (and the TTV) equations are invariant if we measure it from 
              West. This causes no difficulties in the measurements of $\dot{\omega}$, 
              and the East-West discrepancy can be resolved via time-series done via 
              interferometry, astrometry or direct imaging.}              
         \label{FigOmegaDot}
   \end{figure}
%

Eq. (2) is valid if the obliquity of the star is close to zero -- Rossiter-McLaughlin measurements can provide this information.

The planet's contribution to apsidal motion is about 100 times bigger than that of the star (Ragozzine and Wolf 2009). On the l.h.s. the apsidal motion is measurable in principle from TTVs or from RVs as we show hereafter.

The eccentricity is known from RV analysis and/or from occultation observations. RV analysis also provides the mass ratio even for SB1 systems if we have a good estimate for the stellar mass from isochrone fits or from asteroseismology and the secondary's mass is very small with respect to that of the primary. The period and thus the mean motion, the fractional radii $R_i/a$ are known from transit light curve analysis. The rotational period of the star can be estimated accurately enough from its true radius -- provided by isochrone fitting, asterosoismology or by analysis of its emission and distance by e.g. Gaia -- and this, combined with the $V_{e} sini$ value of the star, yields the stellar rotational period.  Out-of-transit light curve modulations can also be utilized to estimate $P_{rot}$ although this technique is limited to stars with significant surface features (spots / plages). Then, $k_2$ of the star can be obtained from theoretical calculations (we use the tables of Claret 2004) and observational results show that such calculations agree reasonably well with the observed values and they vary in a very limited range (Torres et al. 2010). Thus, if we assume a rotational rate for the exoplanet, e.g. that it is synchronous or in a 3:2 resonance (like Mercury) with the orbital period, the Love number of the planet can be determined. We will show that actually the results are not very sensitive to the planetary rotational rate. We note that the argument of periastron can change because of a perturbing third body, as well as by magnetic interaction between the star and planet.

We also note that the Love-number is defined by e.g. Batygin \& Becker (2013) as
\begin{equation} \nonumber
k_{2,Love} =\frac{3 - \eta_2}{2+\eta_2}
\end{equation}
while Sterne (1939) used the so-called apsidal motion constant:
\begin{equation} \nonumber
k_{2,{star~or~planet}} = \frac{3 - \eta_2}{2 (2+\eta_2)} = \frac{1}{2} k_{2,Love}
\end{equation}
where $\eta_2$ is the solution of Radau's equation for $j=2$ at the surface of the star or the planet. The similar notations in previous works are unfortunate, and therefore we use the word 'Love' in the index where we mean the Love-number and we use simple $k_2$ for the apsidal motion constant.


\section{TTV and radial velocity variations under apsidal motion}

We consider the apsidal motion as a linear process in time:
\begin{equation}
\omega = \omega_0 + \dot{\omega} (t - t_0)
\end{equation}
and $\dot{\omega} = \mathrm{constant}$ in time (its second derivative is taken to be zero) where $t$ is the time.

An OXYZ coordinate system is defined in that way that its origin O is in the common center of the mass (CMC), X and Y are oriented to East and North in the tangential plane of the sky, and axis Z connects the observer and the star and it is oriented from the origin away from the observer. The Z-coordinate of the star can be calculated step-by-step as follows:
\begin{equation}
n = \frac{2 \pi}{P}
\end{equation}
$P$ is the anomalistic period (time between two consecutive periapsis passages). This period is constant because it appears in Kepler's third law and we do not consider mass loss from the star and the planet, nor changes in the semi-major axis due to a perturber or tidal forces. (Change in $a$ has much longer time-scale than changes in $\omega$ owing to tidal forces.) Notice that transit observations give the mid-transit time $T_{tr,N}$ at cycle N and that is connected to the anomalistic period as (up to first order in eccentricity, c.f. Gimenez \& Garcia-Pelayo 1983, Eq. (19) and Csizmadia et al. 2009):
\begin{eqnarray}
T_{tr,N} & \approxeq & t_0 + N P \cdot \left( 1 - \dot{\omega} / n \right) + \frac{\vartheta_N P}{2 \pi} \\ \nonumber &-& \frac{eP}{\pi} \cos(\omega_0 + \dot{\omega} (T_{tr,N} - t_0)) \\ \nonumber & + & \mathrm{{\it higher~order~terms~in~eccentricity}} 
\end{eqnarray}
and for the occultation
\begin{eqnarray}
T_{occ,N} & \approxeq & t_0 + N P \cdot \left( 1 - \dot{\omega} / n \right) + P / 2 + \frac{\vartheta_N P}{2 \pi} \\ \nonumber &+& \frac{eP}{\pi} \cos(\omega_0 + \dot{\omega} (T_{occ,N} - t_0)) \\ \nonumber & + & \mathrm{{\it higher~order~terms~in~eccentricity}} 
\end{eqnarray}
One can see that the transit period $P_{tr} = T_{tr,N+1} - T_{tr,N}$ will show oscillations with the apsidal motion period (denoted by U usually, $U = 2 \pi / \dot{\omega}$) and with amplitude of $\sim eP/\pi$ in the first order approximation.

The mean anomaly $M$ of the star is:
\begin{equation}
M(t) = n ( t - t_p ) = n (t - t_0 ) + n(t_0 - t_p)
\end{equation}
where $t_0$ is fixed and $t_p$ is the periastron passage time at the epoch.

Let $t_0$ be the epoch of a well observed transit (in practice, we use a value determined from several transits observed in a short time-window which can be more precise than just a single observation). At that time, the transit occurs when the true anomaly is
\begin{equation}
v_0 = 90^\circ - \omega_0 + \vartheta_0
\end{equation}
$\omega_0$ is the argument of the periastron at the transit epoch. $\vartheta_0$ is a small correction due to the fact that the mid-transit time occurs not at conjuction but at the smallest sky-projected distance of the star and the planet (see Csizmadia 2018, Gimenez \& Garcia-Pelayo 1983, Martynov 1973, Kopal 1959):
\begin{equation}
\tan \vartheta = \mp \frac{e \cos \omega \cos^2 i}{e \sin \omega \pm \sin^2 i \cos \vartheta}
\end{equation}
where the upper and lower signs are valid for the transit and the occultation, respectively, and $i$ denotes the orbital inclination. The eccentric and mean anomalies of the point where the mid-transit  occurred are 
\begin{equation}
\tan \frac{E_0}{2} = \sqrt{\frac{1-e}{1+e}} \tan \frac{v_0}{2}
\end{equation}
\begin{equation}
M_0 = E_0 - e \sin E_0
\end{equation}
$M$ and $M_0$ are actual mean anomalies and its value at the epoch. Then one gets from Eq. (7) that
\begin{equation}
M = M_0 + n (t - t_0)
\end{equation}
The eccentric and true anomalies are:
\begin{equation}
M = E - e \sin E
\end{equation}
\begin{equation}
\tan \frac{v}{2} = \sqrt{\frac{1+e}{1-e}} \tan \frac{E}{2}
\end{equation}
\vspace{1 mm}
\noindent where $E$ and $v$ are the eccentric and true anomalies. If $a_1$ is the semi-major axis of the star around the CMC and $r_1$ its actual distance from CMC, then
\vspace{1 cm}
\begin{equation}
r_1 = \frac{a_1(1-e^2)}{1 + e \cos v}
\end{equation}
\noindent Concerning the $Z$-coordinate:
\begin{equation}
Z = r_1 \sin(v + \omega) \sin i + d + V_{\gamma} t
\end{equation}
where $d$ is the distance to the star from Sun and $V_{\gamma}$ is the systematic radial velocity of the system to the Sun.

%

The radial velocity is the time-derivative of the $Z$-coordinate plus an apparent tidal component, $V_\mathrm{tide}$, not negligible in WASP-18 system:
%
\begin{strip}
\begin{equation}
V_{model} = \dot{Z} + V_{tide} = K \left[e \cos (\omega_0 + \dot{\omega}(t-t_0)) + \cos(v+\omega_0 + \dot{\omega} (t-t_0)) \right] + \frac{\dot{\omega}}{n} \cdot\frac{K  (1-e^2)^{3/2} \cos(v + \omega_0 + \dot{\omega}(t-t_0))}{ 1 + e \cos v}
 + V_\mathrm{tide} (t) + V_{\gamma}
\label{radVelEq}
\end{equation}
\end{strip}
\noindent With $\dot{\omega} = 0$ and $V_\mathrm{tide} = 0$ we get back the usual radial velocity equation free of apsidal motion and ellipsoidal components. We used the usual notation for the RV half-amplitude
\begin{equation}
K = \frac{2 \pi a_1 \sin i }{P \sqrt{1-e^2}}
\end{equation}
while the second term in Eq. (17) comes from the time-derivative of $\omega$. The component denoted by $V_\mathrm{tide}$ stems from the fact that the star is ellipsoidally distorted because of the tidal potential of the planet, therefore we see smaller projected stellar area at conjuctions and larger at quadratures which causes line shape distortions. This is reflected in the observed radial velocities and must be taken into account (Kopal 1959, Arras et al. 2012). A detailed formulation of this effect is given in Section 6.2 by Eqs. (20-26). The model we fit to the observations in Section~5 is described by Eq.~(17).

   \begin{table*}
      \caption[]{Parameters used to calculate the expected apsidal motion rate in Section 4 of WASP-18Ab or used as priors in Eq. (19). The Love number of the star was estimated from the tables of Claret (2004) by interpolation who presented the apsidal motion constant (half of the Love number) as a function of stellar effective temperature, mass and radius. $q$ is calculated from the planetary and stellar masses of Triaud et al. (2010).}
         \label{RadVelSol}
\begin{tabular}{clr}
    \hline
    \hline
Parameter & Value \& Uncertainty & Source \\
\hline
$V_e \sin i$ [km/s] & 11.5$\pm$1.5 & Hellier et al. (2009) \\
\hline
 $R_p / R_s$            & $0.09576^{+0.00105}_{-0.00063}$       & Triaud et al. (2010)   \\
 $R_s / a$              & $0.313^{+0.012}_{-0.009}$           & Triaud et al. (2010)   \\
 $q = M_{planet} / M_s$ & 0.007784$\pm0.000315$     & Triaud et al. (2010)   \\
 $M_s / M_\odot$        & 1.24$\pm0.04$             & Triaud et al. (2010)   \\
 $R_s / R_\odot$        & $1.360^{+0.055}_{-0.041}$            & Triaud et al. (2010)   \\
\hline
 $P_{orb} / P_{rot}$    & $0.17184\pm0.01345$       & calculated from $R_s$ and $V_e \sin i$ \\
\hline
 $a$                    & $0.02047\pm0.00053$       & Southworth et al. (2009) \\
 Inclination [deg]      & $85.96\pm1.70$            & Southworth et al. (2009) \\
\hline
 $T_{eff, star}$ [K]    & $6400\pm100$                  & Hellier et al. (2009)  \\
 Transit Period [d]     & $0.94145299\pm8.7\cdot10^{-7}$ & Hellier et al. (2009)  \\
 Epoch                  & 2~454~221.48~163                 & Hellier et al. (2009), fixed  \\
\hline
$k_{2,star}$            & 0.0143$\pm0.0008$       & Claret (2004)          \\
\hline
$e$                     & 0.0085$\pm0.0020$         & This study   \\
$\omega$ [deg]                    & 257.27$\pm$2.13         & This study   \\
$K$ [m/s]               & $1817^{+1.7}_{-2.0}$      & This study   \\
\hline
\end{tabular}
\end{table*}

\section{Sensitivity analysis}

\subsection{TTVs}

Transit times can presently be determined to a precision of around $20$ seconds with space-based measurements (Csizmadia et al. 2010). This precision is at least a factor of three worse for ground-based observations of shallow transits. Substituting the known system values of WASP-18Ab (see Table~1) and assuming a Love number of WASP-18Ab close to Jupiter's $k_{2,Love} = 0.34$ then we get that the periastron precession rate should be $\dot{\omega} \sim 0.0037$ degrees/day\footnote{A new study by Ni (2018) gives $k_{2,Love} = 0.53$ for Jupiter. This does not change the estimate  significantly.}. By virtue of Eq. (5) we determine that the corresponding TTV-amplitude is about $eP/\pi \sim 3.7$ minutes and the corresponding sinusoidal TTV-curve has a period of $\sim 266$ years. Hence, the annual rate of change in the first ten years is about 0.01 seconds/year. Thus, it is not surprising that the TTV method did not detect it yet. It also predicts that the observed transit period should be highly stable for a long time. This is confirmed by Wilkins et al. (2017) who found that the transit period of WASP-18Ab is stable to less than $6.6\cdot10^{-7}$ days since the discovery (c.f. with the substitution to Eq. (5) which yields much smaller variation in the period than their upper limit). Since TTVs likely do not detect the apsidal motion in this system because the eccentricity is small, the timing precision is low and the observational window is not long enough, we investigate the opportunities provided by RVs.

\subsection{RVs}

Due to the apsidal motion the periastron is shifting and a phase shift will occur which is observable. This shift causes a velocity difference between the actually observed and the one predicted with constant $\omega$, i.e. with $\dot{\omega} = 0$. We illustrate this difference  in Figs. 2-3 where we show the difference in radial velocities between zero and non-zero $\dot{\omega}$-values for different time-periods. Our estimate shows that for WASP-18Ab this should be in the order of 100 m/s difference at its maximum after 2000 days of observations which is about 8 times bigger than the typical observational error bars of one RV-point (Fig. 3). This shows that RV variations are much more sensitive to the apsidal motion than TTVs because they are more accurate.

%
   \begin{figure}
   \centering
   \includegraphics[width=8cm]{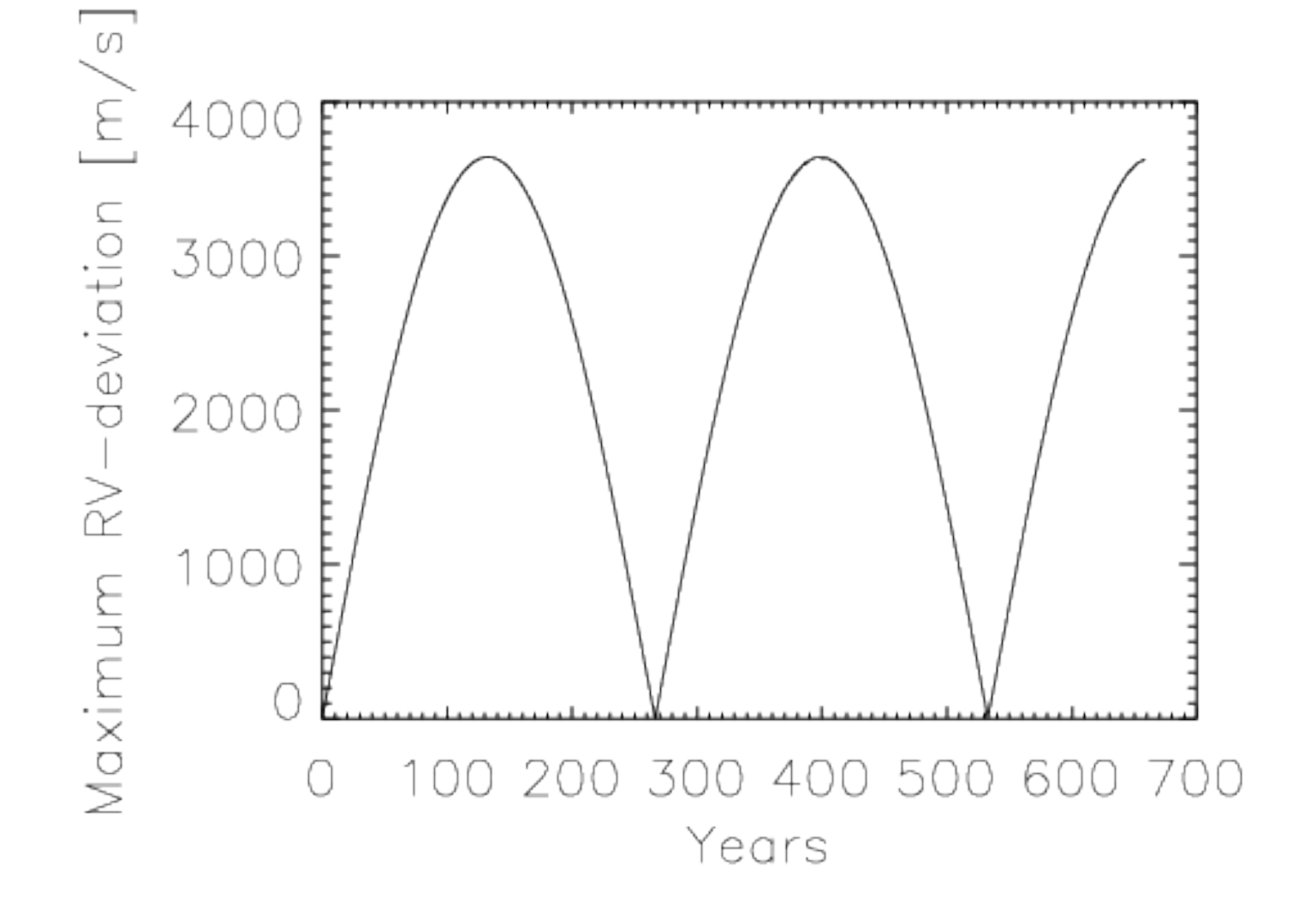}
      \caption{The evolution of the maximum deviation of the 
               observable and the predicted RV-measurements. This maximum 
               value was selected as the maximum deviation over an orbital 
               cycle. The observable RV curve was calculated with the 
               parameters given in Tables~1. 
               The big change is due to the fact that after a while we 
               assume completely opposite phase for the RV-fit if we neglect 
               the apsidal motion.}
         \label{FigRadExp}
   \end{figure}
%

%
   \begin{figure}
   \centering
   \includegraphics[width=8cm]{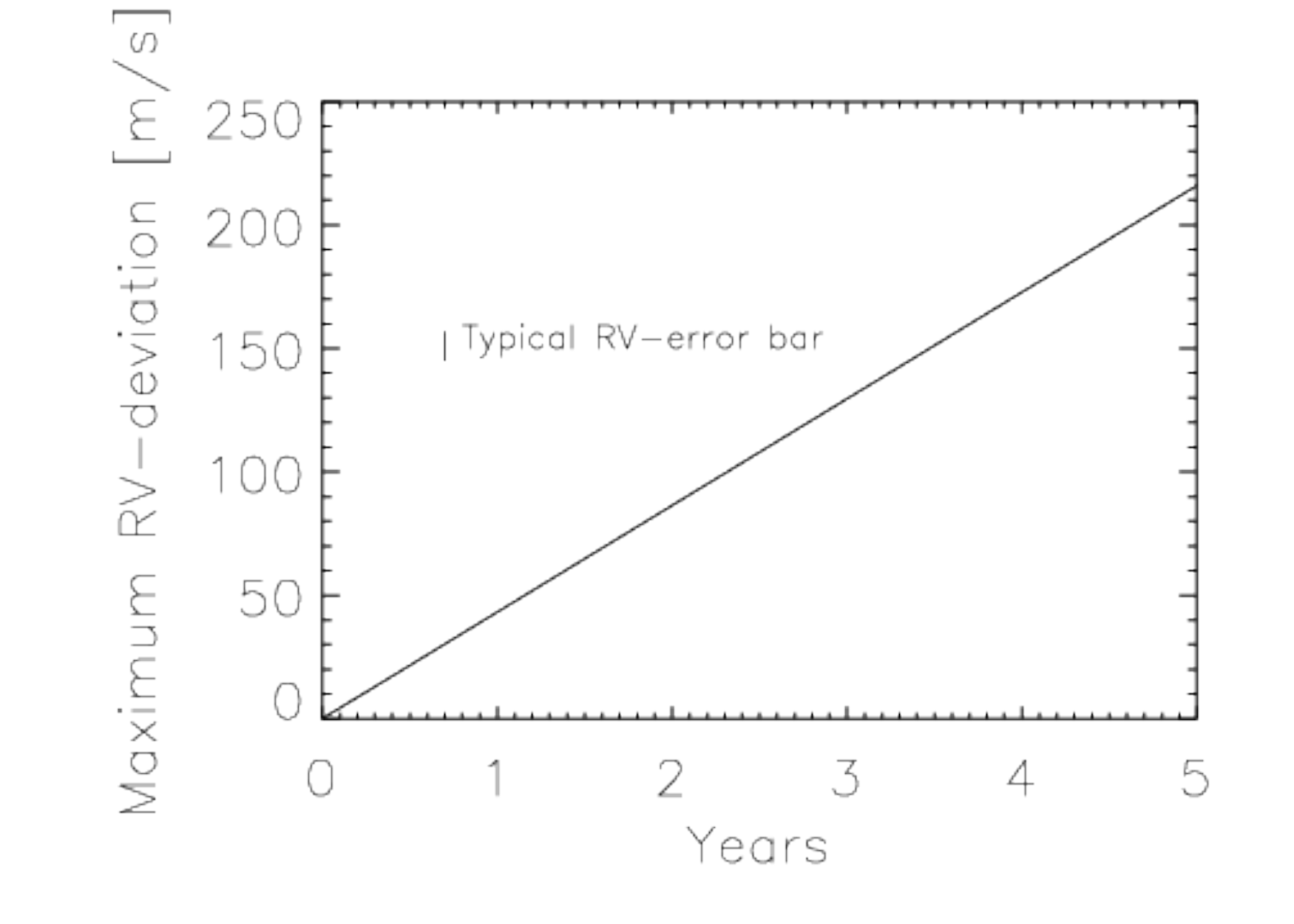}
      \caption{A zoom to the first five years of Figure 2. The typical averaged error 
               bar of the available observations (see Sect. 5) is noted by the thick vertical line. The curve shows the maximum deviation during a revolution from an RV-curve without apsidal motion over time.}
         \label{FigRadExp}
   \end{figure}
%

\section{Data, Model Fit and Results}

We took the 123 published RV observations of Hellier et al. (2009), Triaud et al. (2010), Albrecht et al. (2012) and Knutson et al. (2014). We excluded the points which are between orbital phases $-0.1$...$0.1$, i.e. the ones obtained during primary transit, because we did not fit the Rossiter-Mclaughlin effect. Therefore, we used 54 RV points for the subsequent analysis.

These RV observations span $1849$ days ($=$ $5.06$) years -- long enough to suggest that apsidal motion should be observable in this dataset (Figures 2-3). We carried out four fits with different approaches:

\begin{itemize}

\item[M1:] Model I, we fit $\dot{\omega}$ as a free parameter and 
           we use the RV data points only without using the 
           transit and occultation timing data. We calculated the
           Love number from the derived apsidal motion rate. 

\item[M2:] Model II, we fit $\dot{\omega}$ as a free parameter 
           and 
           we use the RV data points and the transit and 
           occultation timing data simultaneously.  
           This approach allows us to constrain better the 
           argument of periastron and $\dot{\omega}$. As in M1, 
           we calculated the Love number from the measured value 
           of $\dot{\omega}$.

\item[M3:] Model III, we considered the apsidal motion constant (half the Love number) of the planet 
           $k_{2,planet}$ as a free parameter and we fitted Eqs. 
           (2) and (17) together to all data.

\item[M4:] Model IV, where we fixed $\dot{\omega}_{GR} = 0$ and
           $k_{2,star} = k_{2,planet} = 0$ for comparison 
           purposes -- in this model there is no apsidal motion 
           present.

\end{itemize}

\noindent All timings in this paper are in Barycentric Julian Date, in the Barycentric Dynamical Time system ($BJD_{TDB}$).

\subsection{Fit of RV data only (M1)}

In Model 1, we fitted only the RV data. The free parameters were: $V_{\gamma}$, $K$, $\sqrt{e}\sin\omega_0$, $\sqrt{e}\cos\omega_0$ -- to avoid correlations between parameters $e$ and $\omega_0$ (Albrecht et al. 2011) --, $\dot{\omega}$, $P$ and four RV-offset values between the different instruments. $t_0$ was fixed at the epoch given by Hellier et al. (2009). We fitted the model outlined in Sections 3, Eq. (17) by using a Genetic Algorithm-based Harmony Search optimizer (HS, Csizmadia 2018) with 1000 individuals in the population. We tried RV-jitter values of $0$ m/s and $10$ m/s and by a bisection method we adjusted the RV-jitter -- fixed during the HS-run -- until we reached $\chi^2_{RV} = N_\mathrm{obs}-1$, $N_\mathrm{obs}$ being the number of RV-observations.

\begin{table*}
\caption[]{Results of the M1-4 fits.
$P$ is the anomalistic period, not the transit period. Instrument identifiers: 1: PFS, data from Albrecht et al. (2012); 2: CORALIE, data from Hellier et al. (2009), 3: HIRES, data from Knutson et al. (2014), 4: CORALIE, data from Triaud et al. (2010) and we applied a different offset than for the Hellier et al. (2009) data because of possible different instrument settings and data reduction methods; 5: HARPS, data from Triaud et al. (2010). See Section 5.1 for details. The model preferred in Section 6.1  is M3 (RV$+$TTV joint fit, using $k_{2,planet}$ as free fitted parameter instead of $\dot{\omega}$). The prior values used in the fits can be found in Table 1.}
\begin{tabular}{lccccl}
\hline
\hline
Model    &       M1 &   M2 & M3 & M4 & Note \\
Parameter & Value \& Uncertainty  & Value \& Uncertainty  & Value \& Uncertainty& Value \& Uncertainty&  \\
\hline
$V_\gamma$ [m/s]         & $161.9^{+3.1}_{-2.3}$  &                    $163.4^{+2.1}_{-3.9}$  & $162^{+3.6}_{-2.8}$              & $164^{+2.2}_{-3.7}$         \\
$K$ [m/s]                & $1816.3^{+1.9}_{-1.7}$ & $1816.4^{+2.6}_{-1.4}$ &  $1817^{+1.7}_{-2.0}$             & $1816^{+1.7}_{-1.7}$        \\
$\sqrt{e} \sin \omega_0$ & $-0.09295^{+0.00256}_{-0.00549}$  &  $-0.08949^{+0.00394}_{-0.00512}$  & $-0.09013^{+0.00377}_{-0.00462}$ & $-0.09091^{+0.00301}_{-0.00560}$ \\
$\sqrt{e} \cos \omega_0$ & $-0.01169^{+0.00617}_{-0.00617}$  & $-0.01912^{+0.00670}_{-0.00377}$  &  $-0.01872^{+0.00345}_{-0.00345}$ & $-0.00602^{+0.00216}_{-0.00177}$ \\
$\dot{\omega}$ [$^\circ$/day] & $0.0121^{+0.0076}_{-0.0069}$   &       
$0.00907^{+0.00395}_{-0.00177}$   & calculated  & 0 (fixed) & \\
$k_{2,planet}$ & - & - & $0.31^{+0.23}_{-0.10}$ & 0 (fixed)    & aps. mot. const.               \\
RV-offset 2-1 [m/s]      & $3031.9^{+3.7}_{-5.1}$           &  $3032.5^{+3.8}_{-6.7}$           &   $3031^{+3.7}_{-5.4}$             & $3029^{+6.3}_{-3.0}$        \\
RV-offset 3-1 [m/s]      & $ 390.9^{+4.8}_{-4.2}$           &  
$ 393.7^{+4.3}_{-3.1}$           &  $ 395^{+3.8}_{-4.5}$             & $ 393^{+4.7}_{-3.3}$        \\
RV-offset 4-1 [m/s]      & $3166.5^{+3.5}_{-3.0}$           &   $3166.0^{+3.8}_{-2.7}$           &  $3167^{+3.4}_{-3.4}$             & $3167^{+3.7}_{-3.1}$        \\
RV-offset 5-1 [m/s]      & $3180.4^{+4.6}_{-3.0}$           &  
$3181.5^{+4.3}_{-4.3}$           &  $3181^{+4.9}_{-2.9}$             & $3182^{+3.7}_{-4.3}$        \\
$P$ [days]               & $0.94148233_{-0.00001703}^{+0.00001882} $ & $0.94147486_{-0.00000447}^{+0.00001156} $ & $0.94145274_{-0.00000069}^{+0.00000021}$ &  $0.94145251_{-0.00000072}^{+0.00000019}$ &  \\
\hline
Calculated parameters   &                                &   & \\
$e$                     & $0.0088\pm0.0007$              &    $0.0084\pm0.0007$              & $0.0085\pm0.0020$ & $0.0083\pm0.0010$    \\
$\omega_0$ [$^\circ$]   & $262.82\pm3.51$                &    $257.94\pm2.06$                &   $257.27\pm2.13$ & $266.2\pm1.3$      \\
$\dot{\omega}_{GR}$ [$^\circ$/day]    & $0.000707\pm0.000002$ &   $0.000709\pm0.000023$ &  - & 0 & \\ 
$\dot{\omega}_N = \dot{\omega} - \dot{\omega}_{GR}$        & 
$0.01144\pm0.0069$ & $0.00836\pm0.00254$ & - & 0 &  \\
$\dot{\omega}$ [$^\circ$/day] (calculated) & - & - &  $0.0087\pm0.0033 $ & 0 \\
Love-number (=2 $k_2$): \\
$k_{2Love,planet}$          & $0.96\pm0.74$& $0.64\pm0.32$& $0.62^{+0.55}_{-0.19}$ & - & $P_{rot, planet} = P_{orb}$  \\
$k_{2Love,planet}$          & $1.02\pm0.82$& $0.68\pm0.34$& $0.70^{+0.56}_{-0.20}$ & - & $P_{rot, planet} = 0$ \\
$k_{2Love,planet}$          & $0.64\pm0.50$& $0.42\pm0.22$& $0.56^{+0.28}_{-0.20}$ & - & $P_{rot, planet} = P_{orb}/3$ \\
\hline
\end{tabular}
\end{table*}

The results of the HS were refined by running 10 chains of MCMC, with each chain consisting of one hundred million steps. A thinning-factor of 1000 was applied. The marginalized likelihoods were determined and the peak of that distribution defined the final solution. The uncertainties were estimated by calculating the 68\% confidence levels from the marginalized likelihoods.

The correlation plots of the parameters are also produced from the MCMC chains, and we calculated the Pearson correlation coefficients for each parameter pair, as well as the Gelman-Rubin statistic to check possible correlation and convergence problems. The Gelman-Rubin convergence test, denoted by $R$, showed that $R<1.1$ for all parameters, indicating that convergence was reached. 

To get the Love-number, the rotational period of the star can be estimated from the known $R_{star} = 1.36\pm0.06$~$R_\odot$ (Triaud et al. 2010) and from the $V_e \sin i_* = 11.5$ km/s (Hellier et al. 2009), and assuming that stellar rotational axis has $i_* = 90^\circ$ (suggested by Rossiter-Maclaughlin measurements performed by Triaud et al. 2010 and Albrecht et al. 2012 of the sky-projected obliquity of the system which are consistent with zero), we get $P_{orb} / P_{rot} = 0.17184 \pm 0.01345$.

The results of the fit are given in Table 2. The apsidal motion rate - comprising relativistic and Newtonian terms - is $\dot{\omega} = 0.0121^{+0.0076}_{-0.0069}$ degrees/day showing a tentative $1.6\sigma$ detection of apsidal motion based purely on the RV data. For the most likely synchronous planetary rotation case we found $k_{2,planet} = 0.48 \pm 0.37$ and thus $k_{2,Love} = 0.96\pm0.74$. The uncertainties on the system parameters were propagated when we calculated the Love-number from Eq. (2) for different rotational statuses of the planet, listed in Table 2. That is why the uncertainty range of the Love-number is so high.

\subsection{Fit of RV+TTV data with $\dot{\omega}$ (M2)}

In Model 2 we combined the RV-data and the available transit and occultation timing data. Therefore we used the inclination taken from Southworth et al. (2009) to get the value of $\vartheta$ in Eqs. (5-6) for the transit and occultation times. Its error bar was propagated to the fit by drawing the inclination value from a Gaussian distribution.

We took the four mid-occultation times  observed by Spitzer and nine primary transit times observed from the ground (WASP, TRAPPIST, see Triaud et al. 2010, Nymeyer et al. 2011, Maxted et al. 2013) but excluded the less precise amateur measurements published on the ETD page\footnote{{\it http://var2.astro.cz/ETD/}. We make the remark that Zhou et al. (2015) carried out a ground-based $K_s$-band occultation measurement on WASP-18Ab which would extend our baseline, but, unfortunately, they did not publish the mid-time of the occultation, only the transit depth. We also excluded the measurements of Wilkins et al. (2017) obtained by HST because there were long gaps in their transit light curves, decreasing the precision.}. We also did not use the transit timing of McDonald \& Kerins (2018) which is based on sparsely sampled HIPPARCOS data spread over more than three years. Their error bar on the mid-transit time is about 15~min while the expected $O-C$ value of the whole effect we search for is about 3.7~min. We iteratively solve Equations (5, 6, 9) to predict the time of each transit and occultation. By this usage of the known timing of primary and secondary transit events we can constrain our RV-model further, especially the value of the  argument of periastron. TTV and RV data were fitted simultaneously. The results are shown in Table 2. We found $\dot{\omega} = 0.00907^{+0.00395}_{-0.00177}$ degrees/day which is a circa $3.2\sigma$ detection of the apsidal motion. In other words, the inclusion of transit and occultation timing data increased the significance level of the detection because the argument of periastron and $\dot{\omega}$ is better constrained by the timing-data. This yielded $k_{2,planet} = 0.32\pm0.16$, $k_{2,Love} = 0.64\pm0.32$. 

\subsection{Fitting the RV+TTV data with $k_{2,planet}$}

Another possibility to fit the joint RV+TTV data set is to use $k_{2,planet}$ as a free parameter and to calculate the relativistic and the newtonian apsidal motion rates via Eqs. (1-2) for the fit. The advantage of this formalism is to use the posterior distribution for error estimation.

The fitted parameters were in this case: anomalistic period $P$, $V_{\gamma}$, $K$,  $k_{2,planet}$, four RV offsets to take account of the instrumental offsets between the five instruments used, and $\sqrt{e} \sin \omega_0$, $\sqrt{e} \cos \omega_0$. The epoch of transit was taken from Hellier et al. (2009) and fixed. Any uncertainty or error of it was corrected by $\omega_0$. We minimized the $\chi^2$ given by
\begin{eqnarray}
\chi^2 & = & \Sigma_{i=1}^{N_{RV}} \left( \frac{V_{obs,i} - V_{model,i}}{\sigma_i^2 + \sigma_{jitter}^2  } \right)^2 + \Sigma_{j=1}^{N_{T}} \left( \frac{T_{obs,j} - T_{model,j}}{\sigma_j^2  } \right)^2 \\ \nonumber
 & + & \frac{1}{2} \left( \frac{a - a_m}{\sigma_a}\right)^2 \\ \nonumber
 & + & \frac{1}{2} \left( \frac{(P_{orb}/P_{rot}) - (P_{orb}/P_{rot})_m}{\sigma_(P_{orb}/P_{rot})}\right)^2 \\ \nonumber
 & + & \frac{1}{2} \left( \frac{(R_{star}/a) - (R_{star}/a)_m}{\sigma_{R_{star}/a}} \right)^2 \\ \nonumber
 & + & \frac{1}{2} \left( \frac{(R_{planet}/R_{star}) - (R_{planet}/R_{star})_m}{\sigma_{R_{planet}/R_{star}}} \right)^2 \\ \nonumber
 & + & \frac{1}{2} \left( \frac{T_{eff} - T_{eff,m}}{\sigma_T} \right)^2 \\ \nonumber
 & + & \frac{1}{2} \left( \frac{k_{2,star} - k_{2,star,m}}{\sigma_{k_{2,star}}} \right)^2 \\ \nonumber
 & + & \frac{1}{2} \left( \frac{M_{star} - M_{star,m}}{\sigma_M} \right)^2 \\ \nonumber
 & + & \frac{1}{2} \times 10 \times \left( \frac{P_{tr} - P(1-\dot{\omega}/n)}{\sigma(P_{tr})} \right)^2 \nonumber
\end{eqnarray}
where the last eight terms are priors on the semi-major axis $a$, the ratio of the orbital period and the stellar rotation period $P_{orb}/P_{rot}$, the scaled semi-major axis $R_{star}/a$, the radius ratio $R_{planet} / R_{star}$, effective temperature of the star $T_{eff}$, the stellar $k_{2,star}$, the stellar mass $M_{star}$ and observed transit period. The model values (index $m$) were drawn with gaussian distribution centered at their literature values with their one sigma standard deviation (Table~1). We note that they are priors and not fitted parameters, so the model has only 10 free parameters and 8 priors. The priors are used this way to propagate the error bars of the stellar parameters, too.

The last term in the priors expresses that the observed transit period of Hellier et al. (2009) is related to the transit period derived from Eq. (5), neglecting the tiny terms. This decreases slightly the correlation between $k_2$ and anomalistic period.

One can notice that most of the correlations (Fig 4) between the parameters are negligible except three pairs: $k_2$ - period, $k_2$ - $e \cos \omega$ and $e \cos \omega$ - period. This can be understand easily: in the term $\cos (v + \omega)$ of Eq. (17) one can approximate $v = nt + 2 e \sin nt + ...$ and therefore we will have $\cos (v + \omega) = \cos( (n+\dot{\omega})t + \omega_0 + 2 e \sin (nt) + ...)$ and therefore the term $n + \dot{\omega} = 2 \pi / P + 2 \pi /P \cdot k_{2,planet} (R_s/a)^5 \times f(e,q,...) $  causes the correlation between the anomalistic period and $k_{2,planet}$. This correlation can be broken down by more eccentric orbits, more observations and/or longer MCMC-chains. The correlation does not mean that we can measure only the ratio of $k_2/P$ because the anomalistic period $P$ appears in other terms without a relation to $k_2$. As it is clear from Figure 4, the best solutions are concentrated onto a small part of the period-$k_2$ diagram and therefore the correlation just increases the uncertainty of the finally derived parameters but it still allows us to constrain the Love number of the planet.

Because of the five years of the data coverage and of the phase-shifts caused by apsidal motion it is difficult to visualize the fits. Therefore we decided to show the residuals of the fits of the RV, transit and occultation data. These can be seen in Figures 5-7 and the resulting parameters of the fits are listed in Table 2. Notice that the found RV-jitter is in perfect agreement with Albrecht et al. (2012).


We also carried out a fit without any apsidal motion and without apparent tidal RV-term (M4) and the residuals of that fit are compared to the ones of M3 in Figures 5-7 and in Table 4. As one can see, the fit with apsidal motion$+$apparent tidal term provides a better residual curve than without it, preferring M3 model over M4.

%
   \begin{figure*}
   \centering
   \includegraphics[width=18cm]{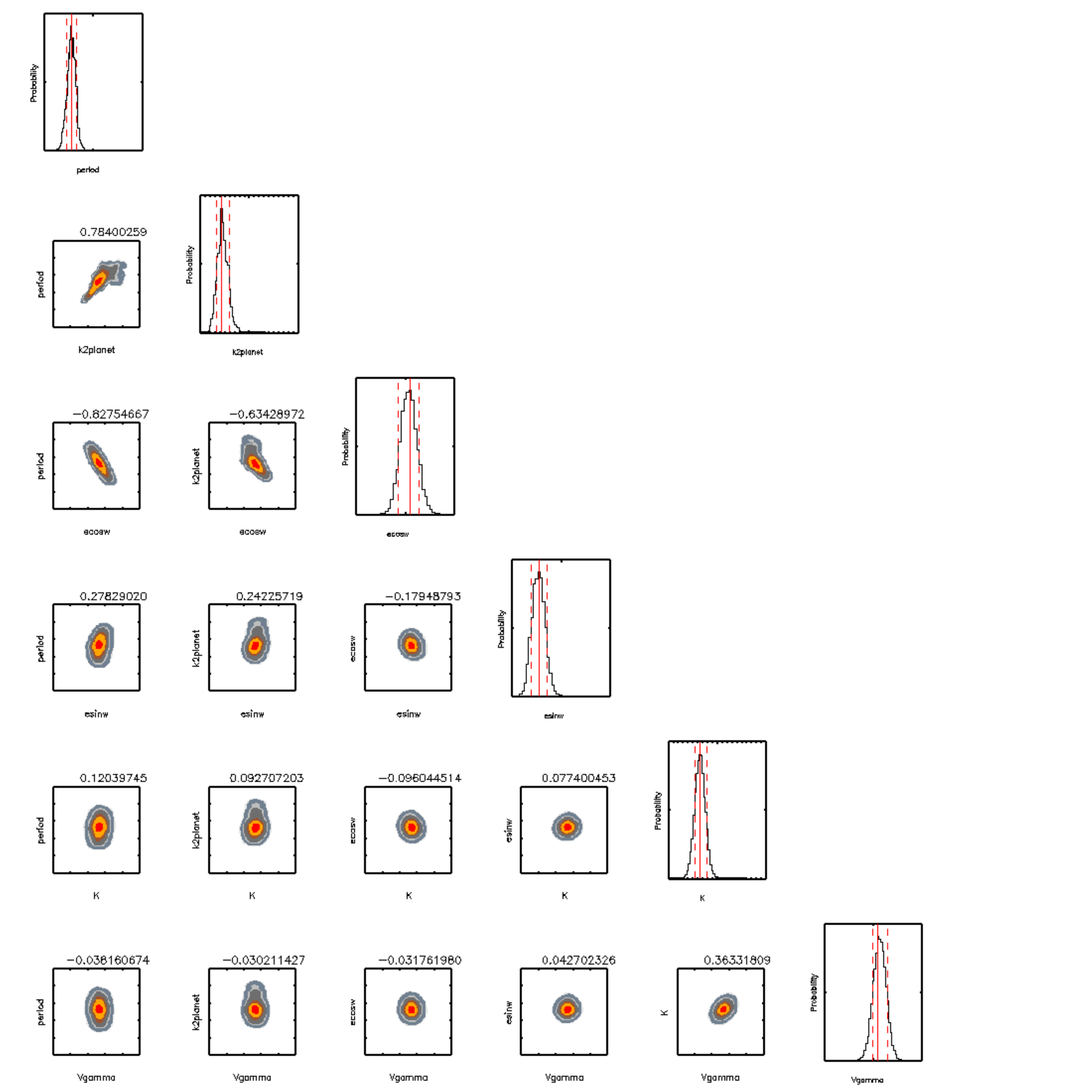}
      \caption{Correlation plots and probability histograms for WASP-18AB RV+TTV fit. Red            
               and orange areas denote the 1sigma region, different gray areas denote the 
               2sigma and 3sigma regions. Numbers over the panels are the values of the 
               Pearson-correlation coefficients between the parameters. The solid and 
               dashed lines show the most probable solutions and the 1sigma limits in the 
               histograms. Only a subset of the parameters are shown.}
         \label{CorrelationPlot}
   \end{figure*}
%

%
   \begin{figure}
   \centering
   \includegraphics[width=8cm]{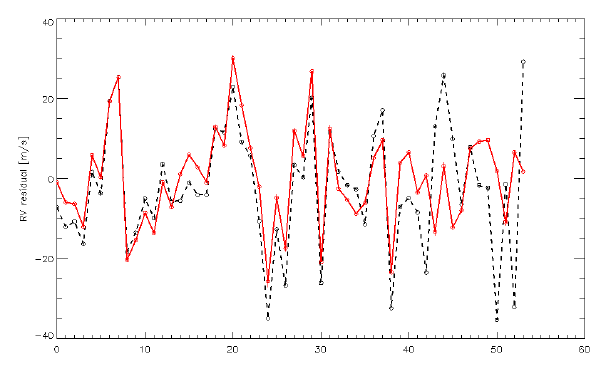}
      \caption{RV-residuals of the model fit. x-axis: the index of 
      the RV points used for the fit, in order of increasing time. 
      y-axis: the residual of the fit in m/s. The dashed black 
      curve represents the residuals of the fit {\it without} any
      apsidal motion because we forced $\dot{\omega}=0$ (including 
      all relativistic and classical tidal terms), as well as 
      $V_{tide}=0.0$ m/s. The solid red curve represents the 
      residuals of the model with apsidal motion (see Eq. (17)). A 
      significant improvement in the residuals can be seen.
      Table 4 helps to identify the observational 
      point via its index.}
         \label{RVresPlot}
   \end{figure}
%

%
   \begin{figure}
   \centering
   \includegraphics[width=8cm]{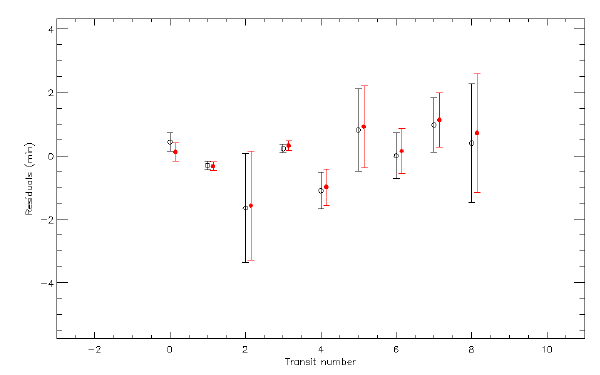}
      \caption{Residuals of the model fit on the transit 
      observations. Notice that the transit observations were 
      taken from the compilation of Maxted et al. (2013). 
      x-axis: the index of the transit observations used for the 
      fit, in order of increasing time (not the cycle number!). 
      y-axis: the residual of the fit in minutes. The open black 
      points represent the residuals of the fit {\it without} any
      apsidal motion because we forced $\dot{\omega}=0$ (including 
      all relativistic and classical tidal terms), as well as 
      $V_{tide}=0.0$ m/s. The red filled circles represent 
      the residuals of the model with apsidal motion (see Eq. (17)). 
      For sake of clarity, the red symbols were shifted horizontally 
      by 0.15 units. Table 4 helps to identify the observational 
      point via its index.
      No improvement can be seen in the transit data with the more 
      complicated apsidal motion model which is contrary to the 
      case of RVs (see Fig. 5.) where we could see significant 
      improvement. This is in accordance with the expectation of 
      Section 3.
       }
         \label{TRresPlot}
   \end{figure}
%

%
   \begin{figure}
   \centering
   \includegraphics[width=8cm]{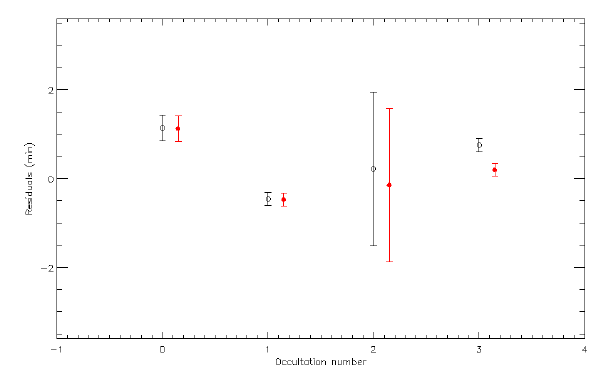}
      \caption{Residuals of the model fit on the occultation  
      observations. Notice that the occultation observations were 
      taken from the compilation of Maxted et al. (2013), too. 
      x-axis: the index of the occultation observations used for the 
      fit, in order of increasing time (not the cycle number!). 
      y-axis: the residual of the fit in minutes. The meaning of 
      symbols is the same as in Fig. 6. Table 4 helps to identify the observational 
      point via its index. A small improvement can be 
      seen in the last occultation measurement with the 
      complicated apsidal motion model relative to the simple 
      model. One can speculate that such an
      improvement might be in accordance with the prediction of Iorio (2011),
      who suggested that a shift of the occultation 
      will be observable within a decade for ultrashort-period 
      planets.}
         \label{OCresPlot}
   \end{figure}
%

\section{Discussion}


Models 2 and 3 give highly consistent values for the Love-number and for the apsidal motion rate, i.e. there is no significant difference between fitting the apsidal motion constant and fitting the apsidal motion itself directly. However, the error propagation in Model 3 is more robust, and so this is our preferred solution. The following discussion is therefore based on the results of Model 3 (Section 5.3, Table 2 and Figures 4-7).

\subsection{Love number of WASP-18Ab}

Assuming a synchronous orbit for the planet $P_{rot,planet} = P_{orb}$, we derive an apsidal motion constant and a Love number for WASP-18Ab: $k_2 = 0.31^{+0.23}_{-0.10}$ and $k_{2,Love} = 0.64^{+0.55}_{-0.19}$. Its significance is $\sim 3.3\sigma$. The apsidal motion rate is about $0.0087\pm0.0033^\circ$/day (Table 2), so its significance means a strong tentative detection at $2.6\sigma$ level. This means an apsidal motion period $U = 360^\circ / \dot{\omega} = 113$ years. When we decreased the error bars of the stellar parameters, the error bar on the apsidal motion rate decreased as well - the biggest contribution to the error bar stems from the scaled semi-major axis. That clearly shows that the uncertainties of the known stellar and system parameters dominates the error bars in the determination of $k_{2,planet}$. Any future improvement in these parameters is  encouraged. We repeated the fit described in Section 5.3 with other rotational rates of the planet, too. If the planet rotational-orbital periods are in 3:2 resonance, then the apsidal motion constant is $k_2 = 0.41_{-0.14}^{+0.23}$. If the planet rotates very slowly ($P_{orb} / P_{rot,planet} =0$) then it is $0.35^{+0.28}_{-0.10}$. In the unlikely event that the planet rotates three times faster than the orbital period, $k_2 = 0.28^{+0.14}_{-0.10}$. (The escape velocity from the planet governs the maximum rotation rate and this means that WASP-18Ab can rotate maximum ca. 24 times faster than its orbital revolution. However, the tidal forces synchronize the orbit quite fast.)

Thus, although some trend can be seen, the planet's rotational speed does not influence the $k_2$-estimation significantly in the realistic cases and within the present level of accuracy. This is because in Eq. (2) the second, tidal term is about fifteen times bigger in case of synchronous rotation than the first, rotational term and therefore it dominates the expression. Such moderate values of $k_2$ are suggestive of a more homogeneous density distribution of a massive hot Jupiter, and the uncertainty range allows us to say that WASP-18Ab has a  similar Love number as Jupiter in our own Solar System (Ni 2018). It is beyond the scope of this paper to establish which giant planet interior models are compatible with the determined Love number.

\subsection{Various views about the eccentricity of WASP-18Ab}

WASP-18Ab was discovered by Hellier et al. (2009) who reported an eccentricity of $e=0.0093\pm0.0029$ based on only nine RV points obtained by CORALIE. According to Eq. (7) of Zakamska et al. (2011), who estimated the expected precision as a function of number of data points, K-amplitude and observational error bars, the uncertainty on the eccentricity in Hellier et al. (2009) is underestimated by a factor of two. However, Zakamska et al. (2011) did not study how the joint fit with transit light curves can increase the precision in eccentricity.

Arras et al. (2012) called attention to the fact that the ellipsoidal shape distortion of the star, caused by the tidal interaction with the planet, introduces a distorted RV signal. They propose that one can observe a small \emph{apparent} eccentricity with an argument of periastron of $270^\circ$ even if the orbit is really circular, and they predicted an RV distortion of $\sim 30 m/s$ for WASP-18Ab. Such spurious, mimicked eccentricities were also mentioned by Kopal (1959, Section V.4). In contrast to Arras et al. (2012) Eq. V.1-22 of Kopal (1959) yields only $\sim 2$ m/s of such RV distortion.

We found that Arras et al. (2012) account only for the effect of tides; they  neglect the effect of stellar rotation, and therefore overestimate the change of shape of the star.  When we set synchronous rotation for the star, then Kopal's formula gave $\sim 24$ m/s of tidally induced RV variations, close to the 30 m/s predicted by Arras et al. (2012). But synchronous rotation is not the case for the host star WASP-18A. Therefore one can propose to revise our views on small exoplanetary eccentricities when it was suspected that they are just apparent eccentricities and stem just from tidal effects. To help such analyses, we give the formula (only the dominating term), based on Kopal (1959) but using modern notation (valid for linear limb darkening with coefficient $u_1$, which is a good enough approximation here) which should be added to the radial velocity model (c.f. Eq. 17):
\begin{equation}
V_{tide} = \frac{K sin(v+\omega) \sqrt{1-e^2}}{\sqrt{1-sin^2 i \cos^2 (v + \omega)}} \frac{P_{orb}}{P_{rot}} \frac{R_{star}}{a} \left( 1 + 1/q\right) \Sigma_{i=2}^{4} f_i w_i
\end{equation}
%
with
\begin{equation}
f_2 = \frac{(8 -  3 u_1)\beta_2 - 5 u_1}{20(3-u_1)}
\end{equation}
\begin{equation}
f_3 = \frac{210+(35-3 u_1) \beta_3 - 50 u_1}{280(3-u_1)}
\end{equation}
\begin{equation}
f_4 = \frac{u_1(\beta_4 +13)}{32(3-u_1)}
\end{equation}
%
and
\begin{equation}
\beta_j = \left( \frac{2j+1}{1+k_{j,star}} + 1 - j \right) \tau_0
\end{equation}
\begin{equation}
\tau_0 = \frac{14388 \mu \cdot K}{4\lambda T_{eff,star} (1-e^{-14388 \mu \cdot K / \lambda / T_{eff,star}})}
\end{equation}
$\mu$ and $K$ stand for micrometer and Kelvin, respectively, and lastly
\begin{equation}
w_j = (1+k_{j,{star}}) q \left( \frac{R_{star}}{a} \right)^{j+1}.
\end{equation}
$q$ is the planet-to-star mass ratio, $\lambda$ is the effective wavelength of the observations (we set $0.6$ micron), $\tau_0$ describes the gravity darkening effect.

Nymeyer et al. (2011) obtained occultation light curves at 3.6 and 5.8 microns and from the phase shift and duration of the transit and the occultation they obtained $e=0.0091\pm0.0012$, a $7\sigma$ significant eccentricity. Maxted et al. (2013) analyzed the 3.6 and 4.5 micron occultation light curves of the system and they reported $e=0.003\pm0.004$ which is compatible both with a circular and the aforementioned eccentric orbits.

Knutson et al. (2014) obtained six new RV points and determined an eccentricity of $0.0068\pm0.0027$ and an argument of periastron of $\omega = 261.1\pm7.4$ degrees. One can speculate that the periastron precession contributes to the error bar in the case of their fixed $\omega$. The significance of their eccentricity is $2.5\sigma$.

Our result: $e = 0.0085\pm0.0020$ is a very strong detection of the eccentricity because we fitted a time-variable $\omega$. Actually, it differs by only $0.2\sigma$ from the eccentricity Hellier's et al. (2009) and it does by only $0.5\sigma$ and $0.3\sigma$ from Knutson et al. (2014) and Nymeyer et al. (2011), respectively\footnote{Here the difference is expressed in term of average sigmas as $\mathrm{difference} = \frac{e_{this~study} - e_{Other~study}}{\sqrt{\sigma_{this~study}^2 + \sigma_{Other~study}^2}}$.}. We consider these differences negligible.

Our $\omega_0 = 257.27^\circ\pm2.13^\circ$ rules out the exact $270^\circ$ argument of the periastron {\it at the epoch} which was fixed at $T_0 = 2~454~221.48163$ (recall that the epoch is taken from Hellier et al. 2009). We must note that the epoch was fixed in our solution and its uncertainty is maybe compensated by $\omega_0$. We suggest obtaining new, very precise, high quality photometric  and radial velocity measurements of WASP-18Ab's primary transit(s) to carry out a new joint fit (c.f. Southworth 2009, 2010).

We propose that the eccentricity is real and it is not caused by tidally induced RV signals. This is supported by substitution to Kopal (1959)'s equation and even more by the Spitzer measurements of Nymeyer et al. (2011). Notice that the observed transit and occultation lengths of Nymeyer et al. (2011) and Hellier et al. (2009) differ by $303 \pm 84$ seconds which is a $3.6\sigma$ difference. Such a difference in the durations is expected for an eccentric configuration.

\subsection{A visual companion to WASP-18A}

We searched the second data release (DR2; Gaia Collaboration et al. 2018) from the Gaia mission (Gaia Collaboration et al. 2016) for close companions to WASP-18A. This revealed an object 11.75 magnitudes fainter than WASP-18A in the Gaia $G$ passband, at a separation of just under 30\arcsec. The two stars have proper motions and parallaxes that are consistent with each other to within 1~$\sigma$ (Table~\ref{companion}). We therefore conclude that they are likely to be a bound pair. Based on the magnitude difference, and the absolute magnitudes tabulated by Pecaut \& Mamajek (2013)\footnote{http://www.pas.rochester.edu/$\sim$emamajek/ EEM\_dwarf\_UBVIJHK\_colors\_Teff.txt}, we suggest that the companion is a late-M type star (approximately M6.5V). Combining the visual separation with the distance to WASP-18A ($123.92 \pm 0.37$~pc), we find that the companion is physically separated from WASP-18A by $3519\pm9$~AU.

Assuming a mass of 0.1~\mbox{$\mathrm{M_\odot}$} for the companion star, we are able to estimate the radial velocity induced by the companion on WASP-18A. When we assume a circular, coplanar orbit with WASP-18Ab then the over five years of observations we can see a maximum velocity change of 
\begin{equation}
V_{max} \approxeq \frac{G(M_{A}+ M_{B}) \Delta t}{a^2_{AB}}
\end{equation}
where $\Delta t = 5.06$ years is the time window between the first and the last RV-observations. A substitution yields $V_{max} \sim 0.1$ m/s radial velocity drift which is negligible in this study. Via Kepler's third law, the orbital period of the stellar companion around the primary star can be estimated to be $\sim 180~000$ years.

We also estimated the impact of the third body on the apsidal motion rate. To do that we used Eq. (12) of Borkovits et al. (2011), assuming a coplanar, circular outer orbit we have that
\begin{equation}
\dot{\omega} \approxeq \frac{3\pi}{2} \frac{M_{B}}{M_{A} + M_{b} + M_{B}} \frac{P_{b}}{P_{B}^2} = 4 \cdot 10^{-15} \mathrm{deg/day}
\end{equation}
which is also negligible.

Because of this companion we changed the planet name from the earlier-used WASP-$18$b to WASP-$18$Ab because it orbits around the brighter primary star, and the faint, newly-discovered companion star is named here WASP-18B.

\begin{table*}
\caption[]{WASP-18A and its stellar companion}
\label{companion}
\begin{tabular}{lcc}
\hline
\hline
Parameter & WASP-18A & WASP-18B \\
\hline
Gaia DR2 ID & 4955371367334610048 & 4931352153572401152 \\
Apparent separation (\arcsec) & 0.0 & $28.398\pm0.001$ \\
Gaia $G$-magnitude & $9.17$ & $20.92$ \\
pmRA (mas yr$^{-1}$) & $25.24\pm0.03$ & $23.65\pm1.98$ \\
pmDec (mas yr$^{-1}$) & $20.60\pm0.03$ & $18.38\pm2.40$ \\
Parallax (mas) & $8.07\pm0.02$ & $9.43\pm1.52$ \\
Distance (pc) & $123.92\pm0.37$ & $106^{+20}_{-15}$ \\
Physical separation (AU) & 0.0 & $3519\pm9$ \\
\hline
\end{tabular}
\end{table*}

\begin{table*}
\caption[]{Radial velocity, transit and occultation timing fit residuals of WASP18Ab system. The index of the observations are the same as in Figures 5-7, respectively. {\it Residual 1} means the residuals without apsidal motion and apparent, tidal-origin radial velocity term (M4 model), {\it Residual 2} represents the residuals of the full model given by Eqs. (2-17) (M3 model). RV-instrument identifiers can be found at Table 2.}
\label{residuals}
\begin{tabular}{lcccccc}
\hline
\hline
RV  & observations & & & & & \\
\hline 
Fig.~5. index & $BJD_{TDB}-2~450~000.0$ & Instr. id. & $RV_{observed} [m/s]$ & Obs. error [m/s] & Residual 1 [m/s] & Residual 2 [m/s] \\
\hline
  0 &          5843.509270 &        1 &              1209.83 &       9.15 &                13.07 &     -13.32 \\
  1 &          5843.728260 &        1 &             -1158.40 &       5.86 &                25.80 &       3.06 \\
  2 &          5843.732260 &        1 &             -1206.06 &       6.65 &                 9.89 &     -12.29 \\
  3 &          5843.837260 &        1 &             -1660.26 &       6.19 &                -6.40 &      -7.93 \\
  4 &          5843.842260 &        1 &             -1645.65 &       6.15 &                 7.92 &       7.58 \\
  5 &          5843.888260 &        1 &             -1557.81 &       6.50 &                -1.66 &       9.27 \\
  6 &          5843.892260 &        1 &             -1542.01 &       6.48 &                -2.28 &       9.64 \\
  7 &          4655.938200 &        2 &              2907.00 &       8.00 &                -7.16 &      -0.92 \\
  8 &          4657.938700 &        2 &              4270.00 &      11.00 &               -10.76 &      -6.35 \\
  9 &          4658.892200 &        2 &              4397.00 &      11.00 &                 1.66 &       5.88 \\
 10 &          4660.935200 &        2 &              4989.00 &       9.00 &                25.34 &      25.38 \\
 11 &          4661.926800 &        2 &              4721.00 &       9.00 &               -13.54 &     -15.42 \\
 12 &          4662.911100 &        2 &              4400.00 &       9.00 &                -5.02 &      -8.56 \\
 13 &          5427.048971 &        3 &              1015.95 &       3.64 &               -23.50 &       0.81 \\
 14 &          6167.071476 &        3 &              1477.27 &       3.87 &               -35.34 &       1.92 \\
 15 &          6193.094044 &        3 &             -1240.09 &       3.82 &                -1.42 &     -10.91 \\
 16 &          6197.032589 &        3 &              -448.66 &       4.08 &               -32.12 &       6.61 \\
 17 &          6209.021271 &        3 &              -839.74 &       4.23 &                29.26 &       1.75 \\
 18 &          4655.938244 &        4 &              3038.59 &       8.35 &               -12.03 &      -5.92 \\
 19 &          4657.938708 &        4 &              4400.45 &      10.57 &               -16.31 &     -12.04 \\
 20 &          4658.892224 &        4 &              4527.77 &      11.16 &                -3.71 &       0.37 \\
 21 &          4660.935178 &        4 &              5119.10 &       9.31 &                19.46 &      19.36 \\
 22 &          4661.926785 &        4 &              4852.26 &       9.19 &               -18.30 &     -20.32 \\
 23 &          4662.911111 &        4 &              4530.95 &       9.18 &                -9.89 &     -13.57 \\
 24 &          4760.700356 &        4 &              5156.68 &       8.51 &                12.42 &      12.96 \\
 25 &          4767.675234 &        4 &              1810.50 &       8.39 &                11.62 &       8.32 \\
 26 &          4767.845516 &        4 &              1821.82 &      11.67 &                22.96 &      30.16 \\
 27 &          4769.805218 &        4 &              2494.59 &      10.15 &                 9.16 &      18.33 \\
 28 &          4770.576633 &        4 &              1521.53 &      14.16 &                 5.75 &       7.65 \\
 29 &          4770.715597 &        4 &              2157.63 &       9.54 &               -10.72 &      -2.02 \\
 30 &          4772.648582 &        4 &              2663.95 &       9.07 &               -35.01 &     -25.67 \\
 31 &          4772.751819 &        4 &              3931.19 &       9.69 &               -12.68 &      -4.78 \\
 32 &          4773.599640 &        4 &              2784.36 &       9.30 &               -26.76 &     -17.38 \\
 33 &          4774.606031 &        4 &              3605.66 &       9.18 &                 3.37 &      12.01 \\
 34 &          4775.655139 &        4 &              4727.74 &       9.66 &                 0.36 &       5.72 \\
 35 &          4776.562493 &        4 &              4446.56 &      10.98 &                20.17 &      26.79 \\
 36 &          4777.543338 &        4 &              4741.72 &      11.46 &               -26.10 &     -20.92 \\
 37 &          4778.581020 &        4 &              5156.85 &       9.02 &                11.64 &      12.46 \\
 38 &          4779.621363 &        4 &              4782.50 &      10.12 &                 1.71 &      -2.55 \\
 39 &          4780.551063 &        4 &              4859.45 &      10.85 &                -1.66 &      -5.35 \\
 40 &          4783.635028 &        4 &              2148.93 &       8.84 &                -2.72 &      -8.84 \\
 41 &          4825.570049 &        4 &              4845.88 &       9.72 &               -11.42 &      -6.20 \\
 42 &          4827.645241 &        4 &              4717.15 &       9.15 &                10.69 &       5.19 \\
 43 &          4831.640624 &        4 &              2247.06 &       8.87 &                17.06 &       9.64 \\
 44 &          4836.591194 &        4 &              1910.00 &       9.75 &               -32.46 &     -23.46 \\
 45 &          4838.557763 &        4 &              2780.64 &      10.05 &                -7.12 &       3.93 \\
 46 &          4854.571845 &        4 &              2894.55 &       9.19 &                -4.91 &       6.58 \\
 47 &          4857.590403 &        4 &              4917.35 &      10.08 &                -8.43 &      -3.44 \\
 48 &          4699.858483 &        5 &              2058.95 &       4.99 &                 3.53 &      -0.96 \\
 49 &          4699.917420 &        5 &              1659.25 &       4.70 &                -5.88 &      -7.13 \\
 50 &          4702.913698 &        5 &              2021.02 &       5.44 &                -5.55 &       1.08 \\
 51 &          4704.818564 &        5 &              2225.01 &       5.28 &                -1.01 &       5.99 \\
 52 &          4706.792686 &        5 &              3258.12 &       5.82 &                -4.03 &       2.78 \\
 53 &          4709.781421 &        5 &              4929.00 &       4.41 &                -4.11 &      -1.01 \\
\hline
\end{tabular}
\end{table*}

\setcounter{table}{3}

\begin{table*}
\caption[]{Cont. of Table 4. Data for transit and occultation observations are taken from Maxted et al. (2013).
References and instrument identifiers: A: Triaud et al. (2010), B: Spitzer, Maxted et al. (2013), 
           C: WASP, Maxted et al. (2013), D: TRAPPIST, Maxted et al. (2013), 
           E: Spitzer, Nymeyer et al. (2011).}
\label{residuals2}
\begin{tabular}{lccccc}
\hline
\hline
Transit & timing & observations & & & \\
\hline
Index in Fig. 6. & $BJD_{TDB}-2~450~000.0$ & Instr. id. & Obs. error [days] & Residual 1 [days] & Residual 2 [days] \\
\hline
  0 &          4664.906250 &        A &             0.0002 &             0.000297 &   0.000080 \\
  1 &          5221.304199 &        B &             0.0001 &            -0.000211 &  -0.000228 \\
  2 &          5419.008301 &        C &             0.0012 &            -0.001146 &  -0.001091 \\
  3 &          5432.189941 &        B &             0.0001 &             0.000159 &   0.000219 \\
  4 &          5470.788574 &        D &             0.0004 &            -0.000762 &  -0.000689 \\
  5 &          5473.614258 &        D &             0.0009 &             0.000564 &   0.000638 \\
  6 &          5554.578613 &        D &             0.0005 &            -0.000000 &   0.000103 \\
  7 &          5570.583984 &        D &             0.0006 &             0.000677 &   0.000787 \\
  8 &          5876.555664 &        D &             0.0013 &             0.000278 &   0.000497 \\

\hline
Occultation & timing & observations & & &  \\
\hline
Index in Fig. 7. & $BJD_{TDB}-2~450~000.0$ & Instr. id. & Obs. error [days] & Residual 1 [days] & Residual 2 [days] \\
\hline
  0 &          4820.716800 &        E &             0.0007 &   0.000792 &             0.000782 \\
  1 &          4824.481500 &        E &             0.0006 &  -0.000319 &            -0.000330 \\
  2 &          5220.833496 &        B &             0.0006 &   0.000153 &            -0.000103 \\
  3 &          5431.719238 &        B &             0.0003 &   0.000524 &             0.000137 \\
\end{tabular}
\end{table*}

\section{Conclusions}

In this study it was shown that there is tentative evidence for the presence of apsidal motion in the WASP-18Ab system. We used archival data spanning more than five years to demonstrate this.

When the origin of this apsidal motion is a combination of general relativistic effects and tidal interaction between the host star and the planet, then it follows that the apsidal motion constant and the fluid Love number of WASP-18Ab are $k_2 = 0.31^{+0.23}_{-0.10}$ and $k_{2,Love} = 0.62^{+0.55}_{-0.19}$, respectively. This result assumes a synchronous rotation for the planet (but not for the star) and the result is thus model and assumption dependent because the stellar fluid Love number is taken from theoretical calculations. We have shown that the Love number is largely insensitive to the actual rotational period of the planet because the tidal contribution to the apsidal motion is about fifteen times larger than that of the rotational modulation.

Such a Love number is compatible with the formerly known Love number of Jupiter in our own Solar System ($k_{2,Love-Jupiter} \sim 0.34$) because of the large error bars, and compatible with $k_{2,Love-Jupiter}=0.53$ proposed by Ni (2018) based on Juno's recent measurements . 

By substitution one can see that about 13\% of the total apsidal motion stems from general relativity and 87\% comes from the tidal interaction. It is not very likely that the observed $\dot{\omega}$ and eccentricity arise from an apparent effect, like that proposed by Arras et al. (2012). If we do not observe any periastron precession, then we must assume that the star and the planet both have zero Love number, or that the orbit is circular against so many findings, or that a third body neutralizes the effect of the apsidal motion, which is particularly unlikely. New RV observations can help to solve this issue.

An unseen third body can also be the reason for the observed apsidal motion, but the lack of TTVs - which is not contradictory to the observed apsidal motion at the found range - and the missing RV trend excludes the presence of a third body with a significant impact on the results. We found a stellar companion to the transit host star, the mutual true separation at the epoch of Gaia observations is about 3519 AU. We estimated its effect on the radial velocity and periastron precession rate and it was found to be negligible. However, this companion may contribute to the excitation of the orbit causing the eccentricity and preventing circularization (e.g. Borkovits et al. 2011).

Thus, the most plausible result is that we can see signs of the apsidal motion generated by tidal interaction and by general relativity in the five year long RV curve of WASP-18Ab -- and the observed Love number also is in the plausible range. We feel, however, that new observations are needed to confirm or reject this result as we are observing a very small signal in a limited dataset.

\begin{acknowledgements}
H. H. and Sz. Cs. thank DFG Research Unit 2440: 'Matter Under Planetary Interior
Conditions: High Pressure, Planetary, and Plasma Physics' for support.
Sz. Cs.  acknowledges support by DFG grants RA 714/14-1 within the DFG
Schwerpunkt SPP 1992, Exploring the Diversity of Extrasolar Planets. SzCs
thanks the  Hungarian  National  Research,  Development  and  Innovation  Office,  for  the  NKFIH-OTKA K113117 and NKFI-KH-130372 grants.
  
This work has made use of data from the European Space Agency (ESA) mission
{\it Gaia} (\url{https://www.cosmos.esa.int/gaia}), processed by the {\it Gaia}
Data Processing and Analysis Consortium (DPAC,
\url{https://www.cosmos.esa.int/web/gaia/dpac/consortium}). Funding for the DPAC
has been provided by national institutions, in particular the institutions
participating in the {\it Gaia} Multilateral Agreement. We thank an anonymous referee for his/her useful comments.

\end{acknowledgements}


\end{document}